\documentclass[aps,prl,10pt,twocolumn,superscriptaddress]{revtex4}
\usepackage{mathrsfs, amssymb, amsmath}  
\usepackage{epsfig, cancel}
\usepackage{latexsym}
\usepackage{natbib, comment}
\usepackage{url}
\usepackage{dcolumn}
\usepackage{multirow}
\usepackage{color}
\usepackage{cancel}
\usepackage{soul}
\usepackage[normalem]{ulem}
\usepackage{amsfonts,amssymb,amsmath, txfonts}
\usepackage{graphicx,epsfig}
\usepackage{psfrag}
\usepackage{hyperref}
\hypersetup{colorlinks=true}
\usepackage{mathtools}
\usepackage{enumitem}
\usepackage{float}
\usepackage[dvipsnames]{xcolor}
\usepackage{xcolor}
\hypersetup{ linktoc=all,
    colorlinks, linkcolor={brightpink},
    citecolor={blue}, urlcolor={blue}
}
\definecolor{rosy}{RGB}{230,235,252}
\definecolor{myframetitle}{RGB}{90,89,170}
\definecolor{myblocktitle}{RGB}{140,185,249}
\definecolor{mytitle}{RGB}{10,80,26}

\definecolor{darkgreen}{RGB}{27,130,45}
\definecolor{darkblue}{rgb}{0,0,0.3}
\definecolor{darkred}{rgb}{0.7,0,0}

\definecolor{light gray}{RGB}{220,220,220}
\definecolor{dark purple}{RGB}{108,0,217}
\definecolor{pink}{RGB}{190,20,100}
\definecolor{orang}{RGB}{193,63,0}
\definecolor{green}{RGB}{11,98,17}
\definecolor{darkpink}{RGB}{153,0,76}
\definecolor{bluegreen}{RGB}{0,102,102}
\definecolor{greenlagan}{RGB}{0,102,0}
\definecolor{redgreen}{RGB}{102,102,0}
\definecolor{Redgreen}{RGB}{153,76,0}
\definecolor{vividviolet}{rgb}{0.62, 0.0, 1.0}
\definecolor{amaranth}{rgb}{0.9, 0.17, 0.31}
\definecolor{palatinateblue}{rgb}{0.15, 0.23, 0.89}
\definecolor{brightpink}{rgb}{1.0, 0.0, 0.5}
\definecolor{cornflowerblue}{rgb}{0.39, 0.58, 0.93}
\definecolor{deepcarminepink}{rgb}{0.94, 0.19, 0.22}
\definecolor{radicalred}{rgb}{1.0, 0.21, 0.37}

\def\H0{{\text{H}\hspace*{-2.05mm}\text{H} 0\hspace*{-1.35mm}0\ }}

\def\be{\begin{equation}}
\def\ee{\end{equation}}
\def\beq{\begin{equation}}
\def\eeq{\end{equation}}
\def\bea{\begin{eqnarray}}
\def\eea{\end{eqnarray}}

\newcommand{\nn}{\nonumber \\}

\begin{document}

\title{Anisotropic Distance Ladder in Pantheon+ Supernovae}

\author{Ruair\'i Mc Conville}\email{ruairi.mcconville@research.atu.ie}
\affiliation{Atlantic Technological University, Ash Lane, Sligo, Ireland}
\author{Eoin \'O Colg\'ain}\email{eoin.ocolgain@atu.ie}
\affiliation{Atlantic Technological University, Ash Lane, Sligo, Ireland}

\begin{abstract}
We decompose Pantheon+ Type Ia supernovae (SN) in hemispheres on the sky finding angular variations up to $4$ km/s/Mpc in the Hubble constant $H_0$ both in the SH0ES redshift range $0.0233 < z < 0.15$ and in extended redshift ranges. The variations are driven largely by variations in absolute magnitude from SN in Cepheid hosts, but are reinforced by SN in the Hubble flow.
$H_0$ is larger in a hemisphere encompassing the CMB dipole direction. The variations we see exceed the errors on the recent SH0ES determination, $H_0 = 73.04 \pm 1.04$ km/s/Mpc, but are not large enough to explain early versus late Universe discrepancies in the Hubble constant. Nevertheless, the Cepheid-SN distance ladder is anisotropic at current precision. The anisotropy may be due to a breakdown in the Cosmological Principle, or mundanely due to a statistical fluctuation in a small sample of SN in Cepheid host galaxies.
\end{abstract}

\maketitle

\section{Introduction} 
The Cosmological Principle (CP), vis-\`a-vis the assumption that the Universe is described by a Friedmann-Lema\^itre-Robertson-Walker (FLRW) spacetime constitutes the bedrock of modern cosmology. Consistency demands that $H_0$ is observationally a constant. Nevertheless, as remarked by Steven Weinberg,  we adopt the CP for a practical reason: \textit{``the real reason for our adherence to the CP is not that it is surely correct, but rather, that it allows us to make use of the extremely limited data provided to cosmology by observational astronomy"}. Within this setting, the cosmic microwave background (CMB) constitutes our most powerful data set. Nevertheless, the amplitude of the dipole ($\ell=1$) temperature anisotropy is orders of magnitude larger than higher order multipoles ($\ell > 1$). As a result, it is interpreted as a feature due to relative motion. Once the dipole is subtracted, this defines the CMB as the rest frame of the Universe, which necessitates correcting observed redshifts for our relative motion.

The CP is traditionally reflected upon whenever large structures are discovered \cite{Gott:2003pf, Clowes:2012pn, Horvath:2014wga, Balazs:2015xsa, Horvath:2015axa, Horvath:2020umh}, most recently a giant arc \cite{Lopez:2022kbz}, or whenever mysterious alignments are found at cosmological scales \cite{Hutsemekers:1998, Hutsemekers:2000fv, Hutsemekers:2005iz, Longo:2011nk, Shamir:2016}. In recent years, the Ellis-Baldwin cosmic dipole test \cite{Ellis1984}, which leverages aberration to infer our velocity, has steadily returned an amplitude excess in the matter dipole relative to CMB expectations \cite{Singal:2011dy, Rubart:2013tx, Tiwari:2015tba, Bengaly:2017slg, Siewert:2020krp, Secrest:2020has} (see also \cite{Tiwari:2022hnf, Dam:2022wwh}). This finding is countered elsewhere \cite{Darling:2022jxt}. However, if substantiated, this suggests that the Universe's isotropic \& homogeneous rest frame either does not exist or has been misidentified. Notably, NVSS radio galaxies \cite{Condon:1998iy} and CatWISE quasars \cite{CatWISE}, observables which are prone to different systematics, now agree on the result \cite{Secrest:2022uvx}.  Nevertheless, since the Ellis-Baldwin test is common to all these studies, theoretical systematics can negate the result \cite{Dalang:2021ruy, Guandalin:2022tyl}. For this reason, one must build a science case beyond the Ellis-Baldwin test and this process has begun \cite{Aluri:2022hzs}. 

Indeed, in the local Universe ($z \lesssim 0.1$) claims of anomalously large coherent peculiar motions, or bulk flows, exist. Most notably, exploiting the kinematic Sunyaev-Zeldovich effect, a ``dark flow" that does not converge to the Hubble flow was reported on scales $\leq 300 h^{-1}$ Mpc \cite{Kashlinsky:2008ut, Kashlinsky:2008us}.  This result is contested by the Planck collaboration \cite{Planck:2013rgv}. More recently, the credence of the dark flow has been boosted by anomalies in galaxy cluster scaling relations at redshifts $z \lesssim 0.1$ that apparently track the CMB dipole direction \cite{Migkas:2020fza, Migkas:2021zdo}. If not due to systematics, one expects confirmation in Type Ia SN. Historically, tests have been performed with SN and anisotropies have been documented in directions aligned \cite{Cooke:2009ws, Colin:2010ds, Li:2013vea, Javanmardi:2015sfa, Colin:2019opb} or closely aligned \cite{Mariano:2012wx, Mariano:2012ia} with the CMB dipole. These directional anomalies become more pressing as the size and statistical power of Type Ia SN samples improve.  

Here we upgrade earlier analysis \cite{Krishnan:2021jmh}, where angular variations of order $\Delta H_0 \sim 1$ km/s/Mpc were noted in the Pantheon sample \cite{Pan-STARRS1:2017jku}, to the Pantheon+ SN sample \cite{Brout:2022vxf, Scolnic:2021amr}. We recover consistent results pointing to a larger $H_0$ in hemispheres encompassing the CMB dipole direction (see also \cite{Krishnan:2021dyb, Luongo:2021nqh}). We find variations in $H_0$ exceeding the errors on the latest SH0ES determination $H_0 = 73.04 \pm 1.04$ km/s/Mpc \cite{Riess:2021jrx} in the same redshift range that are nevertheless too small to fully explain $H_0$ tension. The $H_0$ variations are driven by angular variations in the absolute magnitude of SN in Cepheid hosts, but since $H_0$ in SN cannot be defined without $M$, one concludes that the Cepheid-SN distance ladder is anisotropic at current precision. Our results are consistent with angular $H_0$ variations in SN samples reported elsewhere \cite{Zhai:2022zif, Zhai:2023ubp}. Despite the Pantheon+ sample undergoing some of the most sophisticated redshift corrections for expected and peculiar motions \cite{Carr:2021lcj} to ensure SN are in the Universe's rest frame, $H_0$ remains larger in the CMB dipole direction \cite{Krishnan:2021jmh, Krishnan:2021dyb, Luongo:2021nqh}. Since the local Universe has rich structures, including documented bulk flows towards the Shapley supercluster \cite{Hoffman:2017ako, Howlett:2022len}, angular $H_0$ variations in our cosmic backyard \cite{McClure:2007vv} are expected. In particular, \cite{Wiltshire:2012uh} notes that the Hubble flow is more uniform in local group than CMB frame on scales $\lesssim 150 h^{-1}$ Mpc.

\section{Preliminaries} 
Our analysis follows \cite{Krishnan:2021jmh, Perivolaropoulos:2023iqj, 
Malekjani:2023dky}. Concretely, we truncate the Pantheon+ sample of 1701 data points to a subsample comprising 77 data points in Cepheid host galaxies in the redshift range $0.00122 \leq z \leq 0.01682$ and SN at larger redshifts, i.e. deeper in the Hubble flow. 
We assume the (flat) $\Lambda$CDM cosmology, 
\be
\label{H}
H(z) = H_0 \sqrt{1-\Omega_m + \Omega_m (1+z)^3},  
\ee
where $H_0$ is the Hubble constant and $\Omega_m$ denotes matter density at $z=0$. Next, we define the likelihood, 
\be
\label{likelihood}
\chi^2 = \vec{Q}^{T} \cdot (C_{\textrm{stat+sys}})^{-1} \cdot \vec{Q}, 
\ee
where $C_{\textrm{stat+sys}}$ denotes the Pantheon+ covariance matrix appropriately cropped to include all SN in Cepheid hosts and SN in the Hubble flow in the redshift range of interest. In contrast to \cite{Malekjani:2023dky}, the truncated covariance matrix includes off-diagonal entries correlating SN in Cepheid hosts with SN in the Hubble flow. We further define the vector,
\be
\label{Q}
Q_i = 
\begin{cases} m_i -M - \mu_i,  \quad i \in \textrm{Cepheid hosts,}  \\
m_i - M - \mu_{\textrm{model}}(z_i), \quad \textrm{otherwise.}
\end{cases}
\ee
Here $m_i$ is the apparent magnitude of the SN, $M$ is the absolute magnitude and $\mu_{\textrm{model}}(z)$ is the $\Lambda$CDM distance modulus: 
\bea
\label{model}
\mu_{\textrm{model}}(z) &=& 5 \log \frac{d_{L}(z)}{\textrm{Mpc}} + 25, \nn
d_{L}(z) &=& c (1+z) \int_{0}^{z} \frac{\textrm{d} z^{\prime}}{H(z^{\prime})}.  
\eea
The basic rationale for (\ref{likelihood}) and (\ref{Q}) is that $M$ can be determined from SN in Cepheid host galaxies at lower redshifts $z \lesssim 0.01$, thereby breaking the degeneracy between $M$ and $H_0$ in the model (\ref{model}).  

To test angular variations, following \cite{Krishnan:2021jmh} we convert right ascension ($\mathcal{R}$) and declination ($\mathcal{D}$) angles on the sky into vectors through the identity:  
\be
\label{vec}
\vec{v} = (\cos \mathcal{D} \cos \mathcal{R}, \cos \mathcal{D} \sin \mathcal{R}, \sin \mathcal{D}). 
\ee
We decompose the sky into a grid of points, as illustrated by the smaller black dots in Fig. \ref{fig:dipole_SH0ES_absolute}, and for each dot or direction on the sky, we construct the correponding vector $\vec{v}_{\textrm{sky}}$ from (\ref{vec}). Our grid includes both the point $(\mathcal{R}, \mathcal{D})$ and the antipodal point on the sky, $(\mathcal{R}\pm \pi, -\mathcal{D})$. We next identify the vectors corresponding to all our SN $\vec{v}_i$, before determining inner products, $\vec{v}_{\textrm{sky}} \cdot \vec{v}_i$ and separating SN into hemispheres on the basis of the sign of the inner product. We then construct the likelihood (\ref{likelihood}) for SN in the northern (N) and southern (S) hemispheres, extremise the likelihood to identify the best-fit $(M, H_0, \Omega_m)$ subject to uniform priors, $M \in [-21, -19], H_0 \in [0, 150]$ and $\Omega_m \in [0, 1]$. Finally, we record the absolute difference,  
\be
\label{delta}
\Delta H_0 := H_0^{N}-H_0^{S}, 
\ee 
and an estimate of the statistical significance of the difference, 
\be
\label{sigma}
\sigma := \frac{H_0^N-H_0^S}{\sqrt{(\delta H_0^N)^2+(\delta H_0^S)^2}},  
\ee 
where $\delta H_0^{N,S}$ denote $H^{N, S}_0$ errors. Following \cite{Perivolaropoulos:2023iqj} we estimate our errors from a Fisher matrix, 
\be
F_{ij} = \frac{1}{2} \frac{\partial^2 \chi^2(M, H_0, \Omega_{m})}{\partial p_i \partial p_j}
\ee
where $p_i \in \{M, H_0, \Omega_m \}$. Concretely, we invert the matrix and the errors $(\delta M, \delta H_0, \delta \Omega_m)$ are extracted from the square root of the diagonal entries. Each hemisphere has its own Fisher matrix as we scan over the sky. Once we have performed the scan over the sky, we employ a cubic interpolation of the differences using the python scipy library \textit{scipy.interpolate.griddata} and this interpolation forms the basis of our plots. 

Finally, in both the Pantheon and Pantheon+ samples, evolution of $\Lambda$CDM parameters beyond $z = 0.7$ has been noted \cite{Malekjani:2023dky, Colgain:2022nlb}, so we restrict redshifts below $z=0.7$. This removes high redshift SN, but any reduction in statistical power is small; in the Pantheon+ sample of 1701 SN only 75 are removed. Our objective here is to be conservative and ensure that redshift evolution in $H_0$ does not masquerade as angular evolution. 

\section{Results}
\begin{figure}[htb]
\includegraphics[width=80mm]{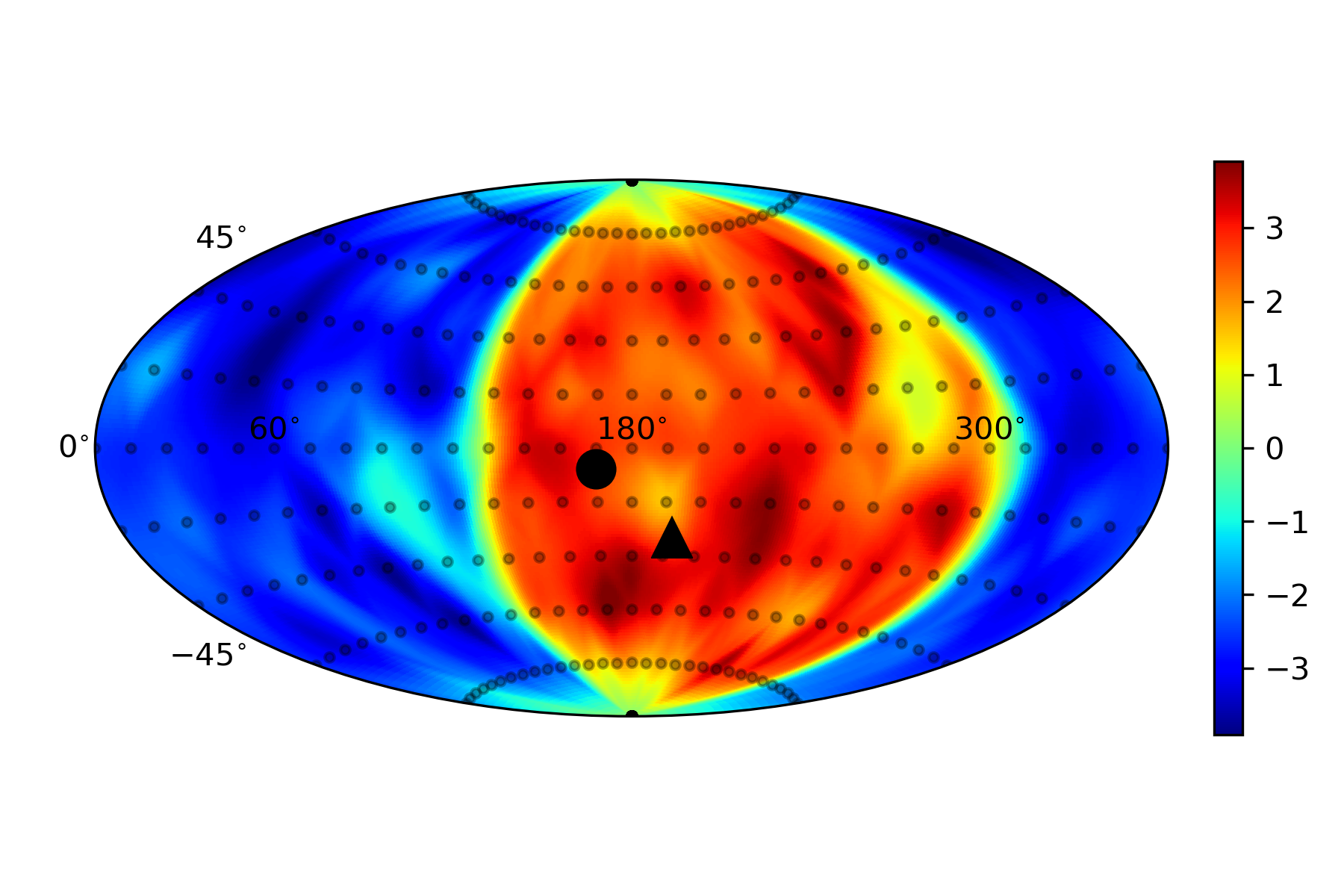} 
\caption{Variations in absolute differences  $\Delta H_0$ (\ref{delta}) over the sky in the SH0ES \cite{Riess:2021jrx} redshift range $0.0233 < z < 0.15$. Small black dots denote sampled points before interpolating. Large black circle and triangle denote the directions of the CMB dipole and the Shapley supercluster.}
\label{fig:dipole_SH0ES_absolute}
\end{figure}

\begin{figure}[htb]
\includegraphics[width=80mm]{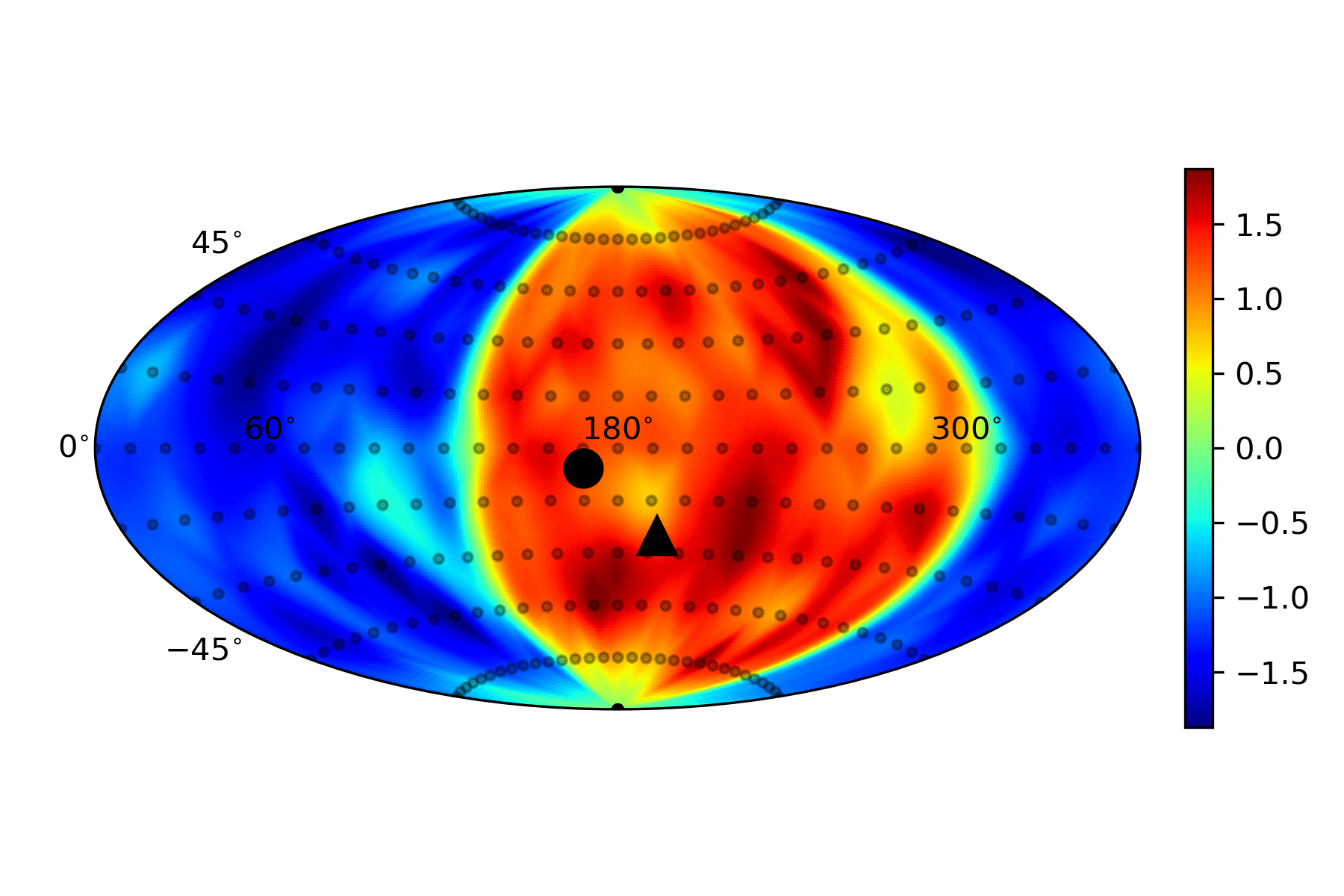} 
\caption{Same as Fig. \ref{fig:dipole_SH0ES_absolute} but $\Delta H_0$ (\ref{delta}) replaced with $\sigma$ (\ref{sigma}).}
\label{fig:dipole_SH0ES_sigma}
\end{figure}

In Fig. \ref{fig:dipole_SH0ES_absolute} and Fig. \ref{fig:dipole_SH0ES_sigma} we show variations in the Hubble constant in the SH0ES \cite{Riess:2021jrx} redshift range $0.0233 < z < 0.15$. From Fisher matrix analysis for the all sky sample in the same redshift range, we have $H_0 = 73.60 \pm 1.10$ km/s/Mpc. In addition, Markov Chain Monte Carlo (MCMC) gives $H_0 = 73.58^{+1.11}_{-1.09}$ km/s/Mpc. Both are in reasonable  agreement with SH0ES, $H_0= 73.04 \pm 1.04$ km/s/Mpc \cite{Riess:2021jrx}. The main point here is that the $H_0$ error is comparable. One notes that $H_0$ is up to  $4$ km/s/Mpc larger in a hemisphere enclosing the CMB dipole direction and the Shapley supercluster. From the points we sample (black dots), the maximum $\Delta H_0 = 3.91$ km/s/Mpc ($\sigma = 1.87$) occurs at $(\mathcal{R}, \mathcal{D}) = (168^{\circ}, -54^{\circ})$. In the CMB dipole direction, $(\mathcal{R}, \mathcal{D}) = (168^{\circ}, -7^{\circ})$, the difference is $\Delta H_0 = 2.62$ km/s/Mpc ($\sigma = 1.18$). It is interesting to consider Shapley because of observations of bulk flows in that direction \cite{Hoffman:2017ako, Howlett:2022len}, which of course may continue beyond. The size of the variation in $H_0$ we see is consistent with earlier findings \cite{Zhai:2022zif}. In Fig. \ref{fig:dipole_SH0ES_sigma} we employ (\ref{sigma}) to estimate the statistical significance of the discrepancy, finding that it is at the $1.5 \sigma$ level at various points on the sky. This is comparable to the statistical significance of $H_0$ variations in the full Pantheon sample \cite{Krishnan:2021jmh}, but Pantheon+ has greatly increased the statistics at low redshifts relative to Pantheon, so the discrepancy is evident at much lower redshifts. Observe that whether we use the $\Lambda$CDM model with two parameters (\ref{H}) or a cosmographic expansion with two parameters in the same redshift range, the result is not expected to change. In other words, we expect the $\Delta H_0$ variations observed in Fig. \ref{fig:dipole_SH0ES_absolute} to be \textit{independent of the cosmological model}.

Note that we are splitting the Cepheid-SN distance ladder on the sky. If one is investigating anisotropies or directional anomalies, it is imperative to perform clean directional splits. Nevertheless, it is possible that the variations we see in $H_0$ are driven more by variations in $M$ from SN in Cepheid hosts and not from SN deeper in the Hubble flow. Since there are only 77 SN in Cepheid hosts, this means that a statistical fluctuation in a small sample \textit{can} explain our results. However, since there are anomalies in other observables in the local Universe \cite{Aluri:2022hzs}, seen holistically, it is prudent to keep physical explanation on the table. In the appendix, we study exclusively SN in Cepheid hosts and document variations in $M$ in the sample. In line with earlier findings \cite{Krishnan:2021jmh, Zhai:2022zif}, we conclude that $M$ variations from Cepheids largely explain the anisotropy we see, but the feature is enhanced by SN deeper in the Hubble flow. {A statistical fluctuation in a small Cepheid host sample could explain the finding, but so too could a Hubble flow that is more uniform in local group than CMB frame at low redshifts \cite{Wiltshire:2012uh}. Low redshift differences disproportionally affect $H_0$ through the lowest rungs of the distance ladder.

In Fig. \ref{fig:dipole_001z07_absolute} and Fig. \ref{fig:dipole_001z07_sigma} we extend the redshift range to $0.01 < z < 0.7$. We find that absolute variations in excess of $\Delta H_0 = 3$ km/s/Mpc persist with comparable statistical significance, but the anisotropies in $H_0$ are visibly less pronounced when higher redshift data $0.15 < z < 0.7$ are included. The maximum variation $\Delta H_0 = 3.08$ km/s/Mpc ($\sigma = 1.60$) occurs at $(\mathcal{R}, \mathcal{D}) = (288^{\circ}, -18^{\circ})$. The difference in the CMB dipole direction is $\Delta H_0 = 1.32$ km/s/Mpc ($\sigma = 0.64$). There may be a simple explanation for $H_0$ variations becoming less pronounced. Of the 800 SN in the range $0.15 < z < 0.7$, 204 are in the same hemisphere as the CMB dipole direction, whereas 596 are not. In contrast, neglecting SN in Cepheid host galaxies, there are 705 SN in the range $0.01 < z < 0.15$, which are split more evenly with 312 SN in the same hemisphere as the CMB dipole and 393 SN in the opposite hemisphere. In short, if there is a difference in hemispheres, it is important to keep an eye on the statistics, as evidently if $H_0$ is lower in the antipodal direction to the CMB dipole, having three times as many SN in one direction than another will increase the weighting for a lower $H_0$ value.  

\begin{figure}[htb]
\includegraphics[width=80mm]{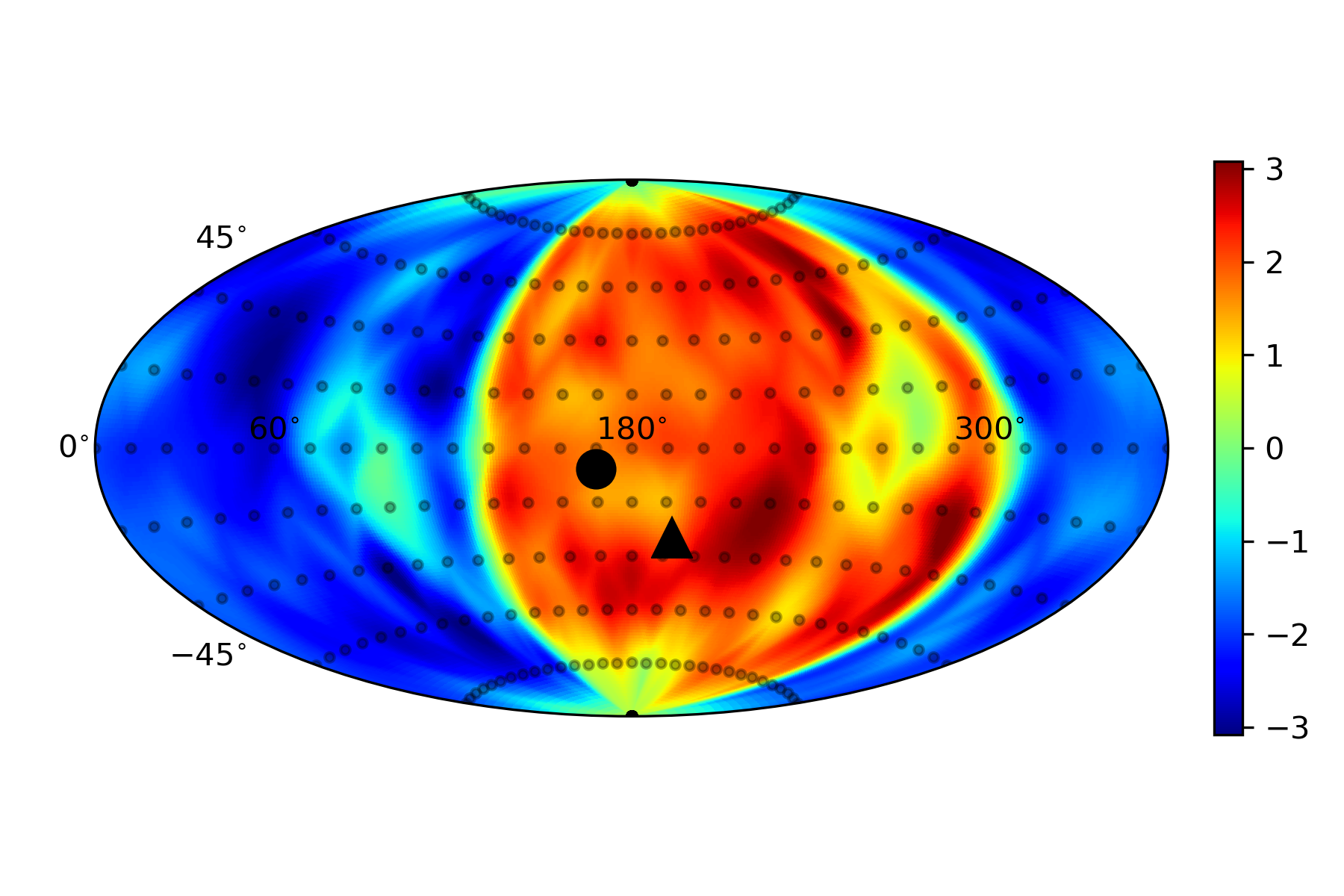} 
\caption{Same as Fig. \ref{fig:dipole_SH0ES_absolute} but with redshift range $0.01 < z < 0.7$.}
\label{fig:dipole_001z07_absolute}
\end{figure}

\begin{figure}[htb]
\includegraphics[width=80mm]{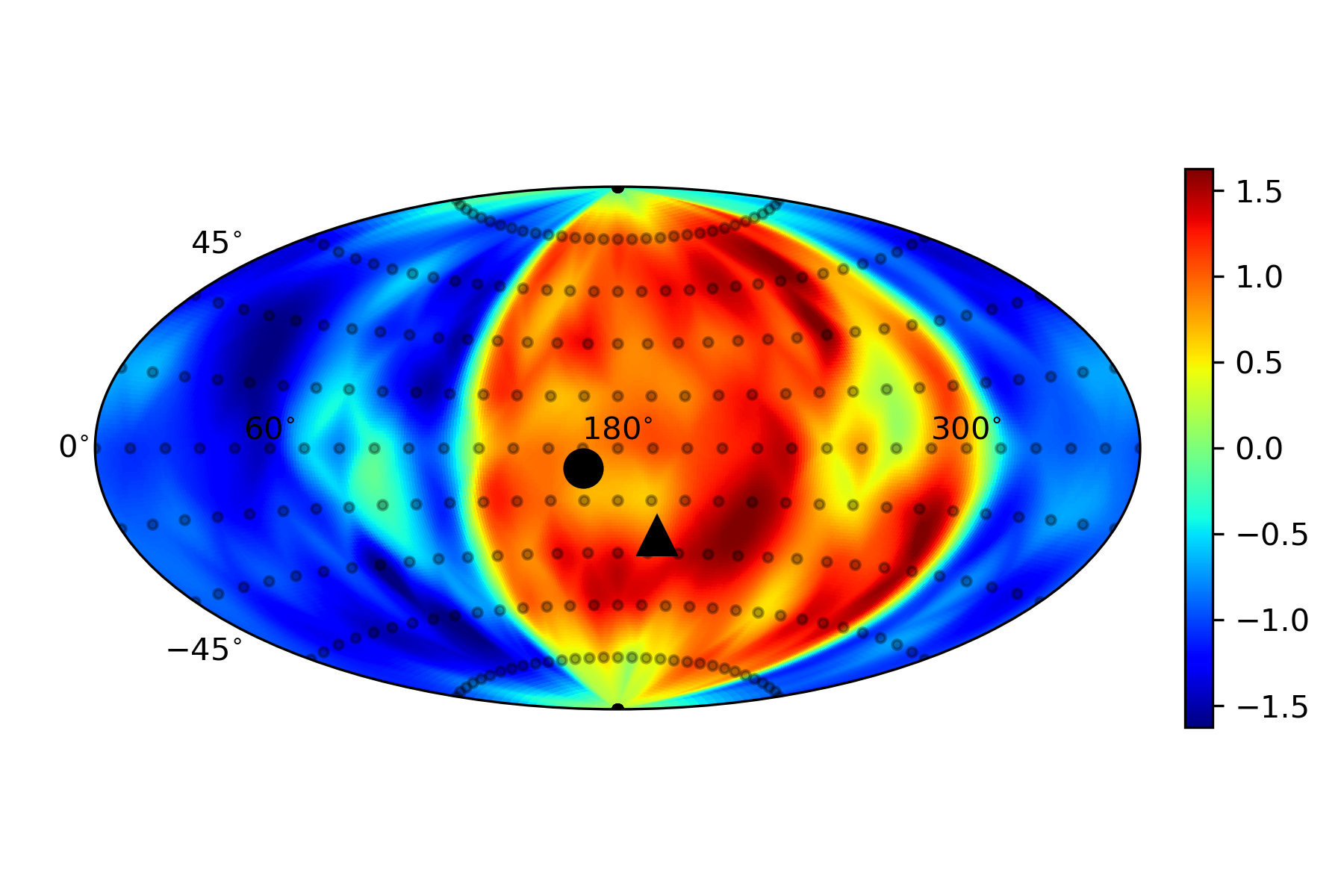} 
\caption{Same as Fig. \ref{fig:dipole_SH0ES_sigma} but with redshift range $0.01 < z < 0.7$.}
\label{fig:dipole_001z07_sigma}
\end{figure}
  
Finally, it was noted in \cite{Krishnan:2021jmh} that the removal of lower redshift $z \lesssim 0.1$ SN had a pronounced effect on the observed anisotropy. One also sees similar variations in the Pantheon+ sample if one fits a dipole \cite{Sorrenti:2022zat}. At low redshifts, the Shapley supercluster can be found in the redshift range $9,000-18,000$ km/s \cite{Quintana:2000vb}, corresponding to $0.03 < z < 0.06$. In Fig. \ref{fig:dipole_006z07_sigma} and Fig. \ref{fig:dipole_01z07_sigma} we show the statistical significance of the $H_0$ discrepancy in the redshift ranges $0.06 < z < 0.7$ and $0.1 < z < 0.7$, respectively. In Fig. \ref{fig:dipole_006z07_sigma}, the maximum $\Delta H_0 = 2.88$ km/s/Mpc ($\sigma=1.42$) occurs at $(\mathcal{R}, \mathcal{D}) = (288^{\circ}, -36^{\circ})$, whereas in Fig. \ref{fig:dipole_01z07_sigma}, the maximum $\Delta H_0 = 3.36$ km/s/Mpc ($\sigma=1.46$) occurs at $(\mathcal{R}, \mathcal{D}) = (144^{\circ}, -18^{\circ})$. In Fig. \ref{fig:dipole_006z07_sigma} the difference in the CMB dipole direction is $\Delta H_0 = 0.81$ km/s/Mpc ($\sigma = 0.36$), whereas in Fig. \ref{fig:dipole_01z07_sigma} the difference is $\Delta H_0 = 0.48$ km/s/Mpc ($\sigma = 0.20$). We omit plots of the absolute differences $\Delta H_0$, as they are similar only with a change of scale, cf. Fig. \ref{fig:dipole_SH0ES_absolute} versus Fig. \ref{fig:dipole_SH0ES_sigma}. Evidently, the feature becomes less pronounced as lower redshift SN are removed, leaving patches on the sky where at most $\sim 1 \sigma$ variations in $H_0$ are evident. This outcome is expected, because as we see from Fig. \ref{fig:SN_above_below_z01}, sky coverage beyond $z=0.1$ is not as good with visibly fewer SN in the same hemisphere as the CMB dipole. Given the poorer sky coverage, it becomes more difficult to sample $H_0$ angular variations in hemispheres on the sky. Nevertheless, one take-home message is that the removal of SN below $z = 0.06$, which could be influenced by the Shapley supercluster, does not fully remove the variations in $H_0$ seen. Note that our sample still includes SN in Cepheid hosts, which exhibit large variations in $M$ (see appendix). Nevertheless, a comparison of Fig. \ref{fig:dipole_006z07_sigma} and Fig. \ref{fig:Msigma} reveals distinct differences in pattern and colouration. Thus, if the variations $ \Delta H_0$ are physical, e. g. due to a local bulk flow, any bulk flow continues beyond Shapley, as the dark flow narrative \cite{Kashlinsky:2008ut, Kashlinsky:2008us} and anomalies in cluster scaling relations \cite{Migkas:2020fza, Migkas:2021zdo} already suggest.

\begin{figure}[htb]
\includegraphics[width=80mm]{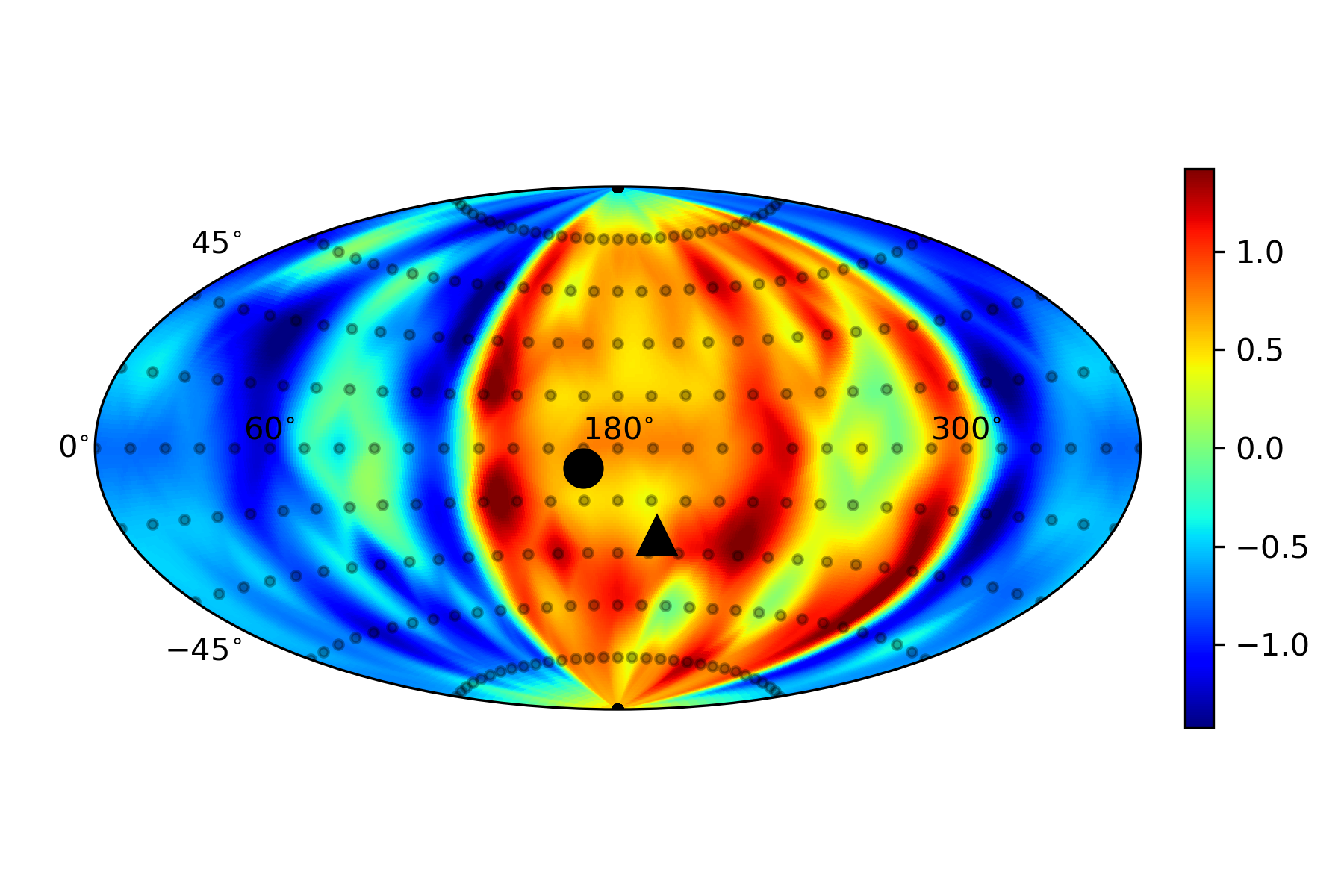} 
\caption{Same as Fig. \ref{fig:dipole_SH0ES_sigma} but with redshift range $0.06 < z < 0.7$, thereby removing the Shapley supercluster.}
\label{fig:dipole_006z07_sigma}
\end{figure}

\begin{figure}[htb]
\includegraphics[width=80mm]{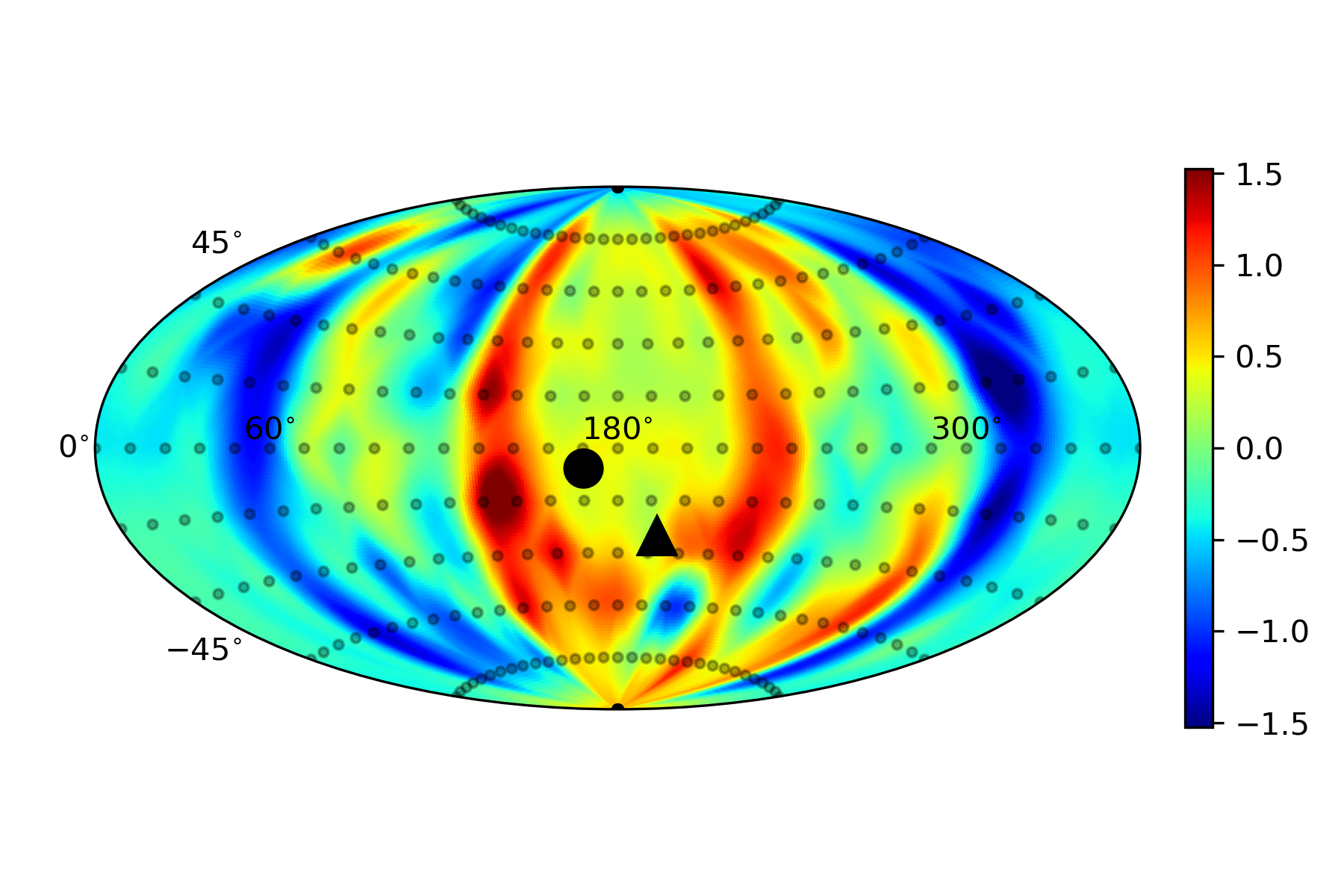} 
\caption{Same as Fig. \ref{fig:dipole_SH0ES_sigma} but with redshift range $0.1 < z < 0.7$.}
\label{fig:dipole_01z07_sigma}
\end{figure}

\section{Discussion}
The Pantheon+ sample \cite{Brout:2022vxf, Scolnic:2021amr} has improved the redshift corrections \cite{Carr:2021lcj} from the Pantheon sample \cite{Pan-STARRS1:2017jku}. In particular, great care has been taken to make sure that SN are in CMB frame, the putative rest frame of the Universe. Nevertheless, as is evident from Fig. \ref{fig:dipole_SH0ES_absolute} - \ref{fig:dipole_001z07_sigma}, variations in the Hubble expansion up to $\Delta H_0 \sim 4$ km/s/Mpc remain, even in the SH0ES range $0.0233 < z < 0.15$. As we explain in the appendix, variations in $M$ in Cepheid host SN contribute to the variation seen in $H_0$, but since one cannot infer $H_0$ directly from SN without a calibrator, the distinction may be moot. Ultimately, our work highlights $H_0$ variations from decomposing the Cepheid-SN distance ladder on the sky. At one end of the spectrum of interpretations, one finds an anisotropic Hubble expansion, e. g. \cite{McClure:2007vv, Wiltshire:2012uh}, at the other, an effect that may be driven by a statistical fluctuation in a small sample of Cepheid host SN. A potential implication is that the SH0ES collaboration are averaging an anisotropic Hubble expansion to get the result $H_0 = 73.04 \pm 1.04$ km/s/Mpc \cite{Riess:2021jrx}. In short, if one is quoting $\sim 1\%$ errors on $H_0$, then the Hubble expansion in the local Universe appears anisotropic at this precision; we may already be looking at a departure from the textbook Hubble-Lema\^{i}tre Law in the Cepheid-SN distance ladder. 

\begin{figure}[htb]
\includegraphics[width=80mm]{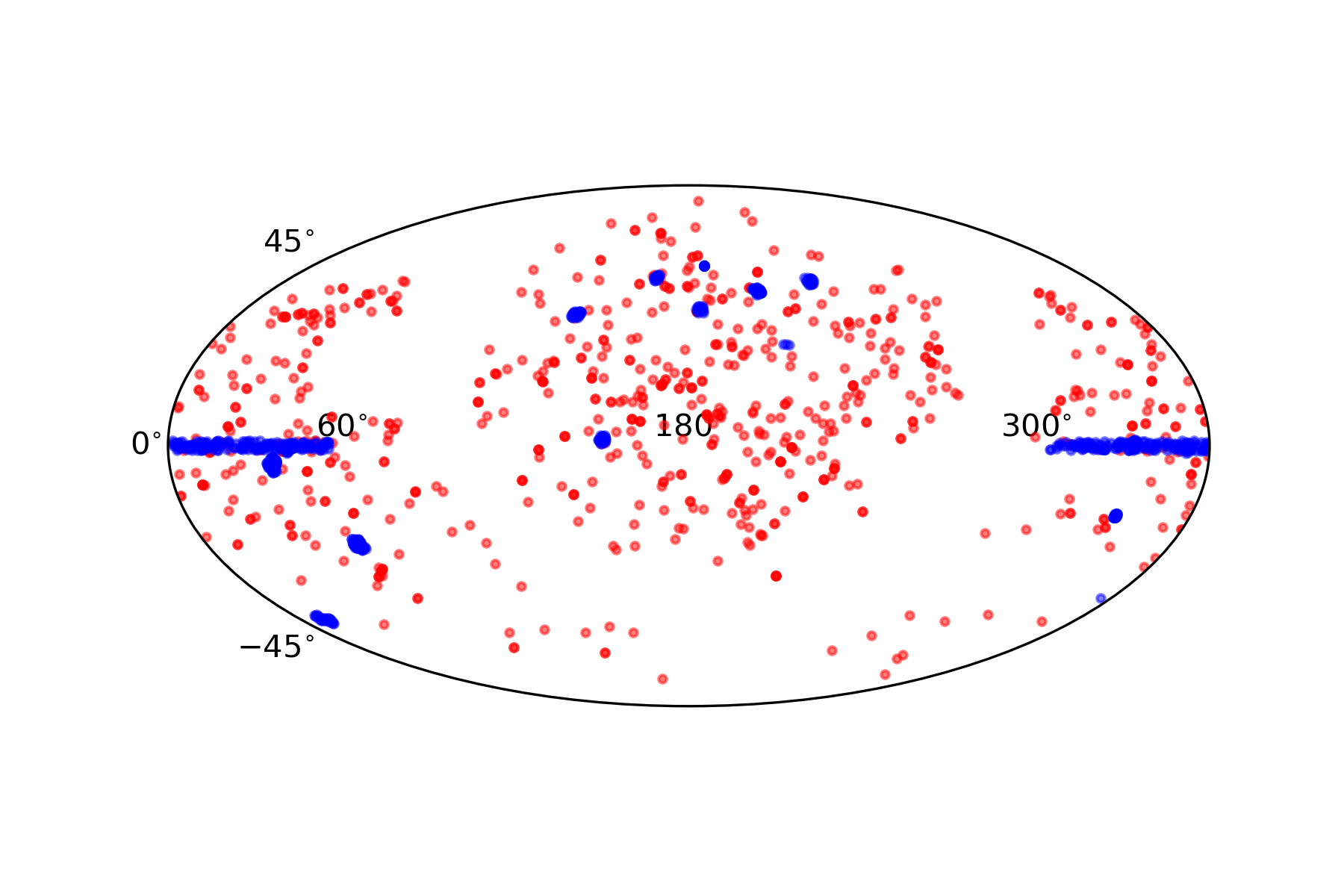} 
\caption{The Pantheon+ sample separated into SN with redshifts above (blue) and below (red) $z = 0.1$. Sky coverage becomes patchy above $z= 0.1$.}
\label{fig:SN_above_below_z01}
\end{figure}

It should be noted that this finding may not be isolated. At the other end of the Universe, two independent studies \cite{Fosalba:2020gls, Yeung:2022smn} have recently highlighted  sizable swings in the $\Lambda$CDM Hubble constant $H_0$ when the CMB is masked. This too seems at odds with final Planck results, $H_0 = 67.4 \pm 0.5$ km/s/Mpc \cite{Planck:2018vyg}, implying that one is averaging over an anisotropy in CMB data. As highlighted in \cite{Aluri:2022hzs}, the well-documented CMB anomaly, the hemispherical power asymmetry (HPA) \cite{Eriksen:2003db, Hansen:2004vq}, could be a plausible explanation for $H_0$ variations in the CMB as the HPA dipole aligns with the axis of the maximum $H_0$ variation. In contrast, in the local Universe, there is a tangible flow in the direction of the Shapley supercluster \cite{Hoffman:2017ako, Howlett:2022len}. Our analysis suggests that lower redshift SN in Cepheid hosts make a definite contribution to the anisotropy in $H_0$, but it persists beyond redshifts associated to Shapley, so Shapley is not expected to explain the feature. Of course, this is consistent with both the original dark flow claim \cite{Kashlinsky:2008ut, Kashlinsky:2008us} and anomalies in galaxy cluster scaling relations \cite{Migkas:2020fza, Migkas:2021zdo}, which both suggest bulk flows out to $\sim 500$ Mpc. In addition, theoretically, FLRW may be unstable to the growth of such anisotropies \cite{Krishnan:2022uar}.   

As highlighted in the introduction, there are now just too many apparent anisotropies in different observables \cite{Aluri:2022hzs} tracking the CMB dipole direction, \textit{despite being in CMB frame}, for this to be a fluke. The only question is whether these anisotropies are due to a local effect, such as an anomalously large bulk flow in the range $z \lesssim 0.1$ \cite{Kashlinsky:2008ut, Kashlinsky:2008us, Migkas:2020fza, Migkas:2021zdo}, or whether they impact cosmology on larger scales. Regardless, no matter how one views it, there is a mismatch between CMB and Pantheon+ SN on the rest frame of a homogeneous \& isotropic Universe, as $\Delta H_0 \sim 4$ km/s/Mpc variations, corresponding to $5 \%$ swings, may be challenging if current precision is $1.4 \%$. The features we see in $H_0$ variations can be modeled through a dipole \cite{Singal:2021crs, Horstmann:2021jjg, Sorrenti:2022zat}, but it is likely the local Universe is more complicated, necessitating a study of higher order multipoles in SN data \cite{Dhawan:2022lze}. 

\section*{Acknowledgements}
We thank Mohammad Malekjani, Saeed Pourojahi and Shahin Sheikh-Jabbari for collaboration on related projects, discussion and comments on the draft. We thank Dillon Brout, Adam Riess and Dan Scolnic for discussion on the physical interpretation of the findings. 

\appendix 

\begin{figure}[htb]
\includegraphics[width=80mm]{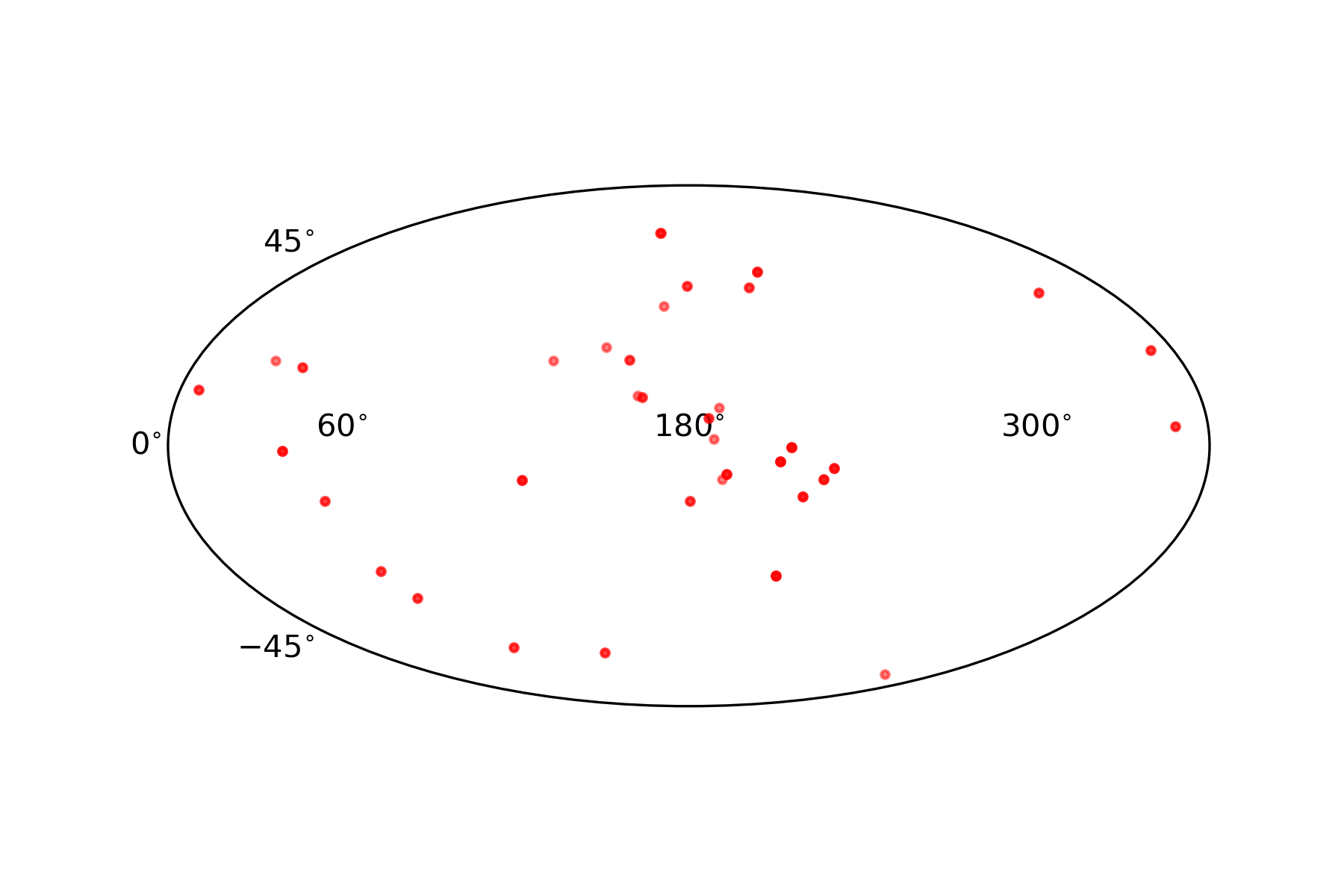} 
\caption{The location on the sky of SN in Cepheid host galaxies.}
\label{fig:Cepheids}
\end{figure}

\section{Variations in $M$ from Cepheid host SN}
In this section we take a look at the 77 SN light curves in Cepheid host galaxies. In Fig. \ref{fig:Cepheids} we document their location on the sky. In Fig. \ref{fig:Mabsolute} we show the variation of $M$, $\Delta M = M^{N}-M^{S}$, where $M^{N,S}$ denote absolute magnitudes in the hemisphere aligned and anti-aligned with the direction of interest on the sky. The statistical significance in Fig. \ref{fig:Msigma} is estimated through $\sigma = (M^N- M^S)/\sqrt{ (\delta M^N)^2 + (\delta M^S)^2}$, where the errors are estimated through a Fisher analysis. Variations in absolute magnitude $M$ up to $0.075$ mag, corresponding to variations up to $1.5 \sigma$, are evident. In a small sample, such variations may be expected. Comparison of Fig. \ref{fig:Mabsolute} and Fig. \ref{fig:Msigma} directly with Fig. \ref{fig:dipole_SH0ES_absolute} and Fig. \ref{fig:dipole_SH0ES_sigma}, respectively, reveals that the variation in $M$ makes a contribution to variations in $H_0$. Nevertheless, the feature is visibly more pronounced in Fig. \ref{fig:dipole_SH0ES_absolute} and Fig. \ref{fig:dipole_SH0ES_sigma}, so there is also a contribution from SN deeper in the Hubble flow. This is in line with findings reported elsewhere \cite{Zhai:2022zif} (see also \cite{Krishnan:2021jmh}).  

\begin{figure}[htb]
\includegraphics[width=80mm]{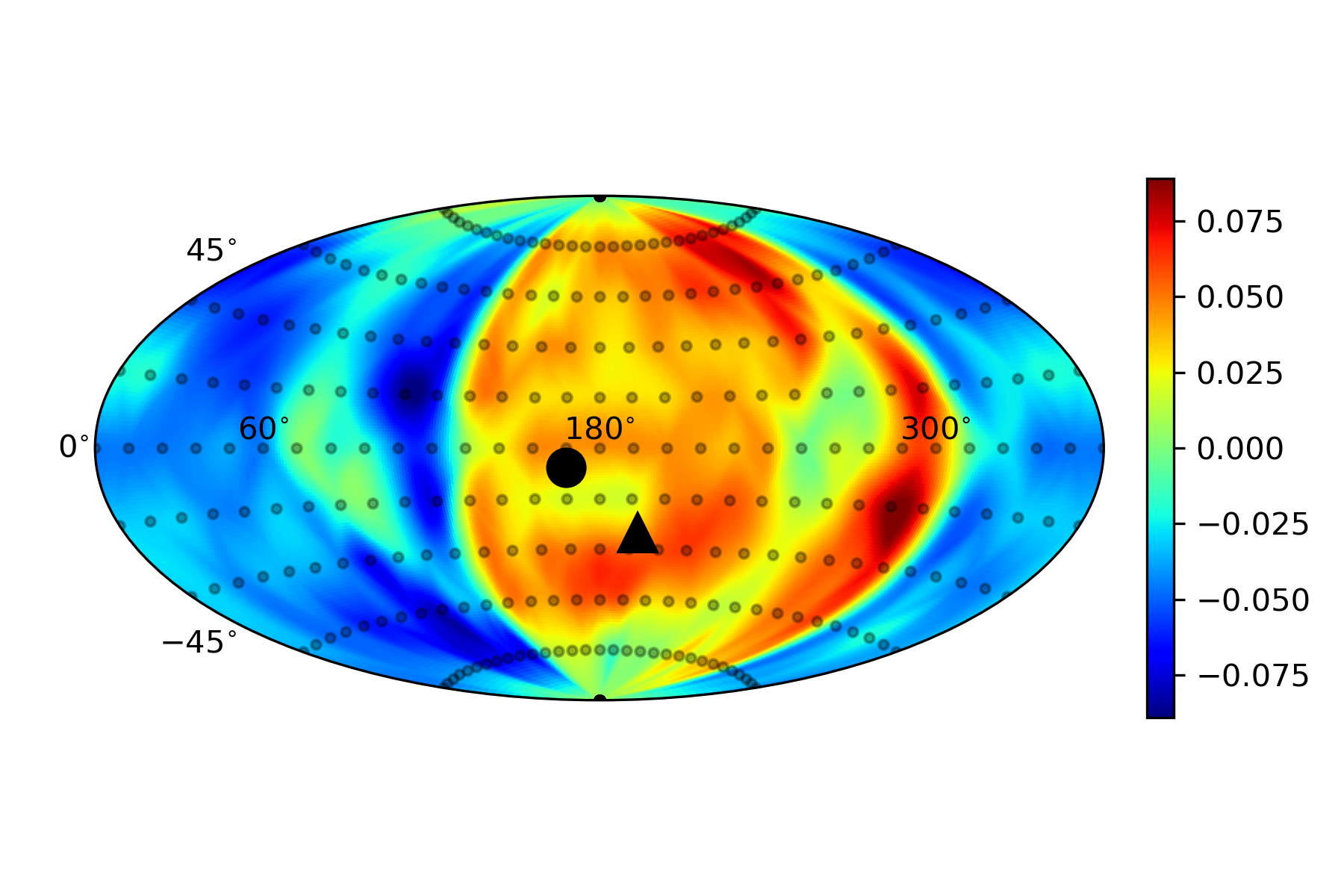} 
\caption{Variations in differences in absolute magnitude $M$ from SN in Cepheid host galaxies on the sky. Small black dots denote sampled points before interpolating. Large black circle and triangle denote the directions of the CMB dipole and the Shapley supercluster.}
\label{fig:Mabsolute}
\end{figure}

\begin{figure}[htb]
\includegraphics[width=80mm]{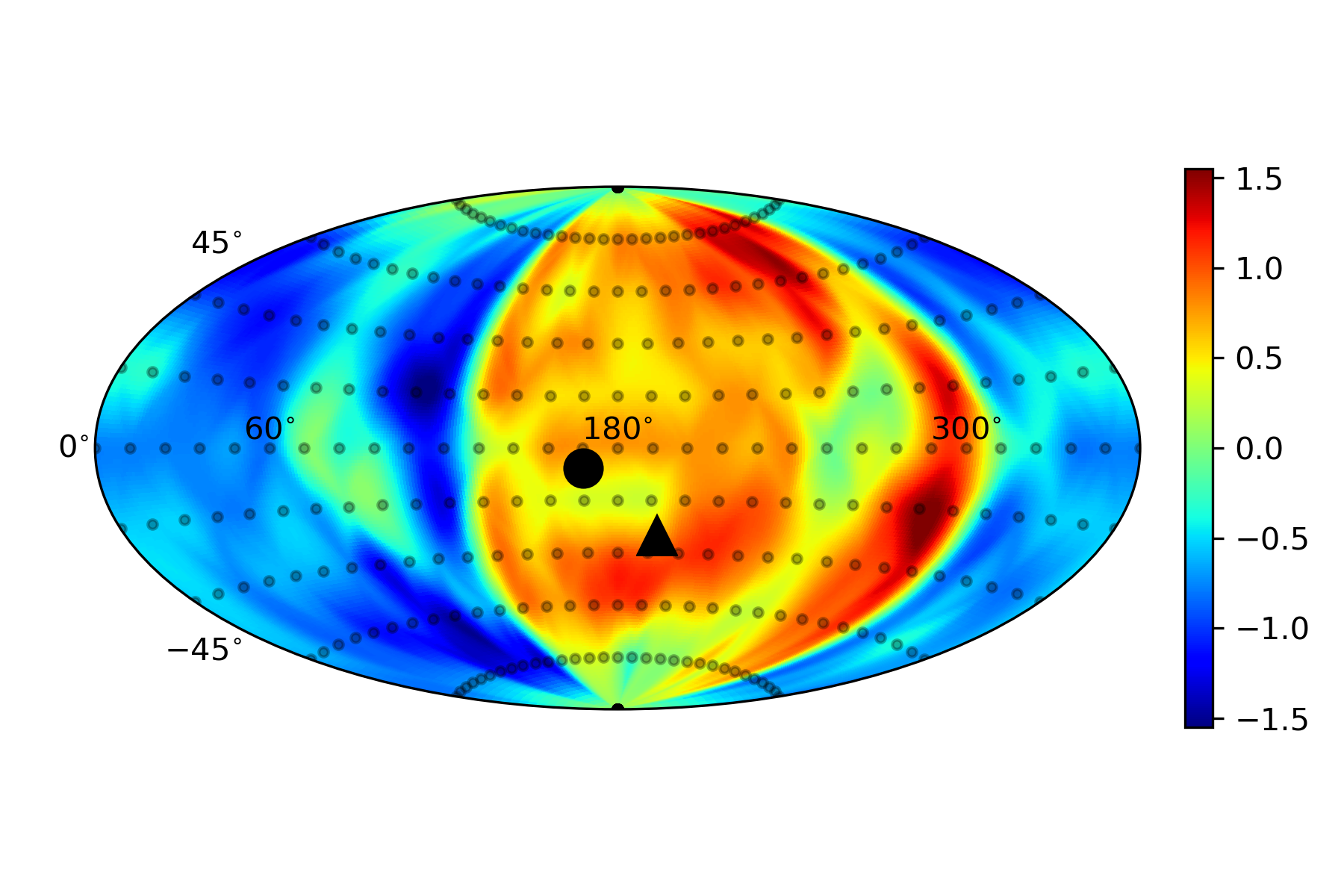} 
\caption{Same as Fig. \ref{fig:Mabsolute} but showing an estimate of the statistical significance of absolute magnitude $M$ variations.}
\label{fig:Msigma}
\end{figure}

\end{document}